\documentclass[amssymb,twocolumn,
prl,aps,floatfix,showpacs]{revtex4}
\usepackage{graphicx}  % needed for figures
\usepackage{bm}        % for math
\usepackage{amssymb}   % for math
\usepackage{amsmath}
\usepackage{verbatim}   
\usepackage{subfigure}
\usepackage{upgreek}
\newcommand{\intensity}[2]{$#1\times10^{#2}\mbox{ Wcm}^{-2}$}
\newcommand{\fs}[1]{#1~fs}
\newcommand{\as}[1]{#1~as}
\newcommand{\um}[1]{#1~$\upmu$m}

\begin{document}
\title{XUV-initiated high harmonic generation: driving inner valence electrons using
below-threshold-energy XUV light}
\author{A. C. Brown} 
\author{H. W. van der Hart}
\email{andrew.brown@qub.ac.uk}
\homepage{}
\affiliation{Centre for Theoretical Atomic, Molecular and Optical Physics,
School of Mathematics and Physics, Queen's University Belfast, Belfast BT7 1NN,
United Kingdom}

\date{\today}

\begin{abstract}

  We propose a novel scheme for resolving the contribution of inner- and
  outer-valence electrons in XUV-initiated high-harmonic generation in neon.  By
  probing the atom with a low energy (below the $2s$ ionisation threshold) ultrashort XUV pulse, the
  $2p$ electron is steered away from the core, while the $2s$ electron is
  enabled to describe recollision trajectories. By selectively suppressing the
  $2p$ recollision trajectories we can resolve the contribution of the $2s$
  electron to the high-harmonic spectrum. We apply the classical trajectory
  model to account for the contribution of the $2s$ electron, which allows for
  an intuitive understanding of the process.

%  We investigate XUV-initiated high-harmonic generation in neon to assess 
%  the contribution of the inner-valence $2s$ electron to the harmonic yield. The $2s$ contribution extends the cutoff energy of the spectrum by
%  30~eV and by using ultrashort (250~as) XUV pulses to ionize from the $2s$ shell, it is
%  possible to suppress the contribution of the $2p$ electrons. We employ a simple extension to
%  the three-step model to handle trajectories initiated by
%  photoionization rather than tunnel ionization, which allows an intuitive understanding
%  of the process.
\end{abstract}

\maketitle

%\section{Introduction}

High harmonic generation (HHG) is now a well established tool both for the
generation of high energy and attosecond laser pulses
\cite{67as_pulse,attosecond_pulse_train}, and as a measurement technique for
atomic and molecular structure and ultrafast electron dynamics
\cite{multielectron_atoms, multielectron_molecules, tomographic_molecular}. The
well-known `three-step' model captures well the gross dynamics
of the process: an electron is 1) ionized then 2) driven by a strong laser field
before 3) recolliding with its parent ion, and releasing its energy in the form
of a high-harmonic photon \cite{corkum1993}. 

For some time now, researchers have been concerned with both the {\it
optimization} of the process \cite{mixed_gas_efficiency, as_train_efficiency,
improved_efficiency_hhg}, and its {\it application} to measurement in so called
high-harmonic spectroscopy \cite{attosecond_spectroscopy_review,
hhg_spectroscopy_valence, multielectron_atoms, multielectron_molecules,
tomographic_molecular}. Both of these considerations raise the same 
conclusion: the initial step of the three-step model (tunnel ionization) limits
both the conversion efficiency and the ability of HHG to probe general electron
dynamics, restricted as it is to the emission of the outermost valence electron.
Hence, it has been proposed that by subjecting the target to both the strong
driving field and a short XUV pulse, the HHG process can be initiated by
photoionization rather than tunnel ionization \cite{as_train_efficiency}. This
has the advantage of being significantly more efficient, and of opening the
possibility of driving more deeply bound electrons. 

Several theoretical studies
have explored the efficiency question by utilizing simple model calculations to
describe the three-step process \cite{xuv_hhg_biegert,xuv_hhg_gademan}. A
related scheme-- wherein an XUV pulse excites a core electron to a
valence hole during the valence electron's transit-- has also been studied
\cite{xuv_hhg_buth_theory,xuv_hhg_buth_krypton,
xuv_hhg_tudorovskya}. More
recently, an 
XUV-{\it initated} HHG (XIHHG-- i.e. the core, rather than the valence, electron is
emitted in the initial step)  scheme was used to elucidate core-hole dynamics in small molecules
\cite{correlation_core_hole_dynamics}. 

The attraction of studying {\it core}-hole dynamics is two-fold.  Firstly, the
HHG cutoff is extended in proportion to the electron binding energy, and thus by
driving HHG with a deeply
bound electron substantial extension of the HHG plateau can be achieved.  Not
only does this allow the generation of ever shorter, attosecond pulses, an
extended cutoff also allows greater scope for analysis of the spectrum.
Secondly, high energy photoionization favors ionization of core- rather than
valence-electrons, whereas the outer electrons are more susceptible to tunnel
ionization and photoionization by lower energy XUV light. Thus, to a reasonable
approximation, only the core electron responds to the XUV photon, and the
response of the outer electrons can be neglected.

However, one can envisage dynamics involving the correlated motion of
inner- and outer-valence electrons which may be probed with XIHHG. For
instance, the window resonances in the photoionization spectrum of argon are
due to the interference of $3s$ and $3p$ electrons. We have previously shown that
these interferences also impact on low energy harmonic generation in
short-wavelength (390~nm) fields \cite{brown_prl}. To
investigate their role in {\it high} harmonic generation, driving wavelengths in the
near- to mid-IR range should be employed, but at such low photon energies the
response of the inner-valence
electron is negligible. Thus a combination of XUV and IR light might be
used to probe such interference dynamics.

There are only a handful of computational methods capable of
describing the multielectron dynamics necessary for a thorough exploration 
of HHG with inner-valence electrons \cite{TDCIS,
TDRASSCF-multielectron_atoms}. Among them, the R-matrix with time-dependence
(RMT) technique has demonstrated significant promise for the description of
ultrafast processes in general multielectron systems \cite{RMT}. RMT has been applied
variously to HHG in the computationally challenging near-IR wavelength regime
\cite{ola_nearIR}, electron rescattering in negative ions
\cite{ola_rescattering}, IR-assisted ultrafast ionization of positive ions
\cite{RMT_rydberg} and attosecond transient absorption
spectroscopy of doubly- and core-excited states \cite{RMT_ATAS}. 

In the present manuscript we explore the XIHHG process in neon and assess the
contribution of the inner-valence electron to the resulting spectra. In contrast
with previous studies, which promote the inner-valence electron during the
transit of the outer valence electron \cite{xuv_hhg_buth_theory,
xuv_hhg_buth_krypton, xuv_hhg_tudorovskya}, we drive the inner-valence electron
on a three-step-like trajectory by employing XUV pulse-energies {\it below} the
$2s$ ionization threshold. Additionally, we consider how the timing of the
ultrashort XUV pulse can be used to selectively enhance the contribution of the
$2s$ electron, an extension of the method previously explored by Schafer {\it
et~al.} \cite{path_control_efficiency}.

%\section{Theory}
The RMT method is an {\it ab-initio} technique for solving the time-dependent
Schr\"{odinger} equation for general, multielectron atoms or ions in strong
laser fields. We employ the well-known R-matrix paradigm, whereby the interaction space is split into
two regions: an inner region-- close to the nucleus, wherein full account is
taken of all multielectron interactions-- and an outer region-- where a single,
ionized electron moves under the influence of the laser field and the long-range
potential of the core. Several implementations of time-dependent R-matrix theory
exist \cite{guan_tdrm,tdrm,analytical_r_matrix1,analytical_r_matrix2}, but RMT offers
the most robust and general numerical approach for tackling processes in
strong-fields. This is achieved through representing the wavefunction with a B-spline based, R-matrix
basis in the inner region, and with a highly efficient grid based approach in
the outer region \cite{RMT}.

The neon target used for this study is discussed in
detail elsewhere \cite{window_resonances_argon}. Briefly: the neon
atom is described in a close-coupling with pseudostates scheme as a
Ne$^+$ ion plus an electron.  The descriptions include all
$2s^22p^5\epsilon\ell$ and $2s2p^6\epsilon\ell$ channels up to $L_{max}=139$.
The inner (outer) region boundary is 32 a.u. (2000 a.u.), 
(results are converged with respect to these boundaries). The set of
continuum orbitals contains 70 B-splines for each angular momentum of the
continuum electron. 

The primary laser pulse used is a 3-cycle (\fs{18}), \um{1.8} pulse with an
intensity of \intensity {1.2}{14}. The XUV intensity is 10\% of the IR
intensity. Both the IR and XUV pulses have a $\sin^2$ ramp on/off profile.
Unless otherwise stated, the delay between IR and XUV pulses is chosen to be
\fs{2.75} such that the XUV peak arrives one half-cycle before the IR peak.
According to the classical three-step model, electrons `born' into the field at
this time describe the optimal trajectories for HHG. 

Calculations performed for various XUV intensities (1\% and
0.1\% of the IR intensity), show
that the overall yield scales linearly with the XUV intensity.
Calculations performed with 800~nm IR pulses instead of 
\um{1.8} show that the mechanisms discussed below also apply at the lower
wavelength. We here display only the results for the \um{1.8} IR and 10\%
intensity XUV as they support longer plateaus and clearer observation of the
key features. The probability to tunnel ionize from the $2p$ shell in the IR field
is $0.00003\%$, and to photoionize with the 45~eV XUV from the $2p$ or $2s$ is $0.3\%$ and
$0.002\%$ respectively.

%\section{Results}
\begin{figure}[t]
\begin{centering}
\includegraphics[width=7.5cm]{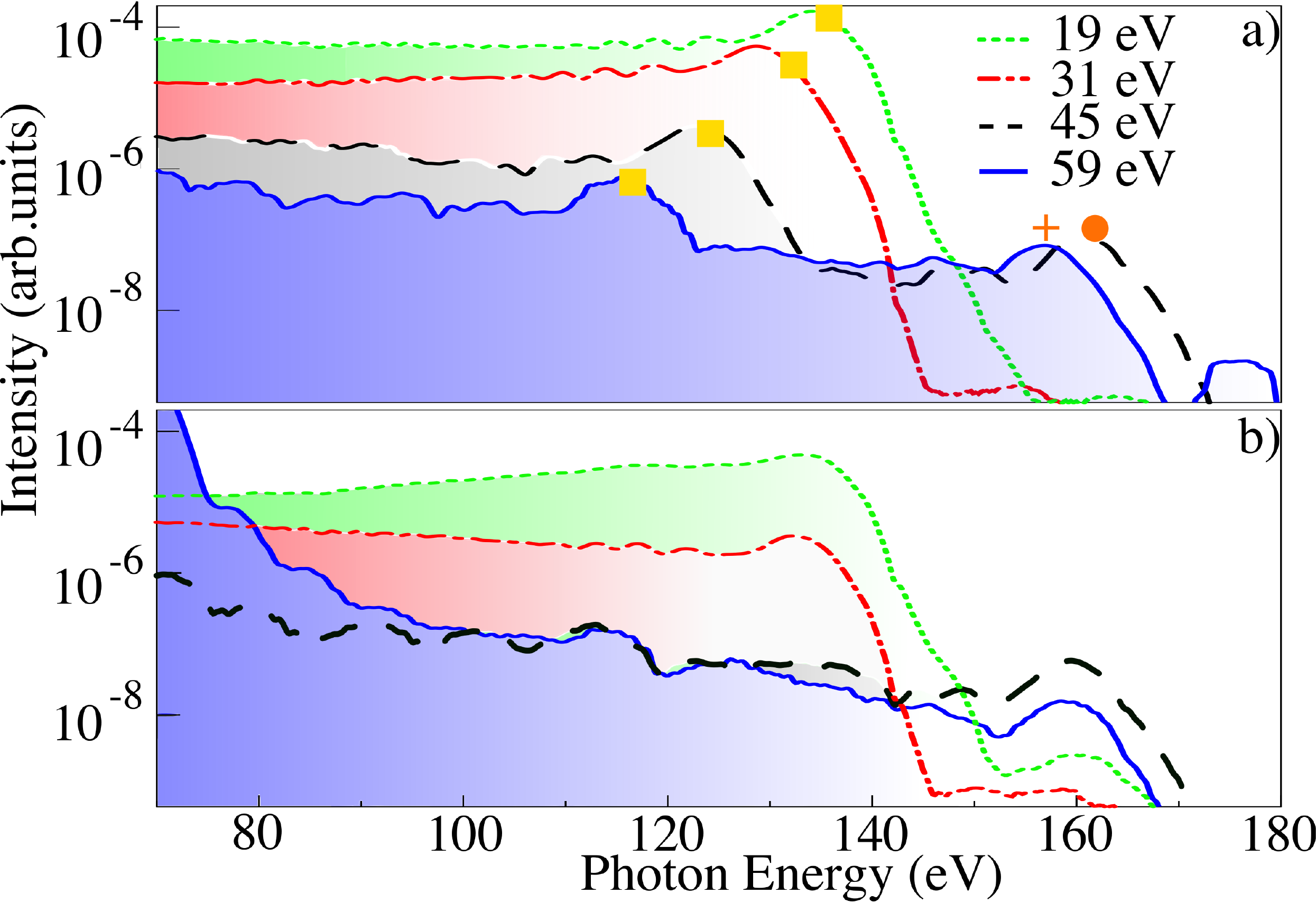}
\caption{(Color online) The smoothed harmonic yield from Ne driven by an \fs{18},
\um{1.8}, \intensity{1.2}{14} pulse, in
combination with a a) \fs{1} and b) \as{250} XUV pulse of energy 19~eV (black
dotted line), 31~eV (red dot-dashed line), 45~eV (green dashed line), and 59~eV
(blue solid line). The cutoff of the $2p$ harmonics as predicted by the model
(see text) are marked by yellow squares. The predicted $2s$ cutoff 
is marked for the 45~eV (circle) and 59~eV (`+') XUV
pulses.
  \label{fig:short_long}}
\end{centering}
\end{figure}

We perform calculations
for XUV-photon energies ranging from beneath the $2p$ binding energy ($E_{2p}=
21.6$~eV) to above
the $2s$ binding energy ($E_{2s}=48.5$~eV), and use two XUV pulse durations: \fs{1} (`long') and
\as{250} (`short').
Figure \ref{fig:short_long} shows the
harmonic spectra for four XUV-photon energies-- 19, 31, 45, 59~eV (corresponding
to particular odd harmonics of the \um{1.8} pulse) in both the long- and
short-pulse configurations. 

% rephrase this paragraph to highlight the importance of the below threshold
% photon
At a photon energy of 19~eV (45~eV), the XUV pulse is not sufficiently
energetic to ionize the $2p$ ($2s$) electron directly. However, the HHG yield is
still increased by several orders of magnitude over the IR-only spectrum
(not shown).  We attribute this to a two-photon
(IR+XUV) ionization process. These XUV photon
energies are resonant with a number of Rydberg states converging onto the $2p$
and $2s$
ionization thresholds: and thus photo-excitation to a (superposition of several) Rydberg
state(s) is followed by field ionization. This two stage
process allows the electron to be `born' into the field
with approximately zero energy, which is crucial for optimizing the three-step
recollision process. 

%By initiating the HHG
%process with an XUV pulse, the harmonic yield has been increased by between 4
%and 7 orders of magnitude relative to the IR-only spectrum
%(not shown). The
%largest increase in yield is observed for an XUV photon energy of 19~eV, which
%is below the ionization potential of Ne ($I_p = 21.6$~eV). Hence, the process
%cannot initiated by direct photoionization of the $2p$ electron, but rather a
%two-photon (IR+XUV) process. The 19~eV photon
%is resonant with a number of $2s^22p^5{}ns$ states, and
%we suggest that the $2p$ electron is ionized in a two stage process: excitation to
%a (superposition of several) Rydberg state(s), and subsequent field ionization.
%Calculations for time delays 
%between 2.75 and \fs{3.25} (not shown) show the same enhancement in yield, corroborating the proposed two stage process.

For XUV photon energies of 45~eV and 59~eV Fig. \ref{fig:short_long} shows an extension
of the plateau, the cutoff of which is dependent on the XUV photon
energy. In the long-pulse case there appears a double plateau
structure and the cutoff energy of the first
plateau (marked by yellow squares) decreases
with increasing photon energy.

To understand the extension of the cutoff we consider the empirical formula
for the cutoff energy \cite{cutoff_law}:
\begin{equation*}
  E_{\mathrm{cutoff,2p}} \approx 3.2U_p+I_p =  137~\mathrm{eV}, 
\end{equation*}
where $U_p$ is the ponderomotive energy, and $I_p$ the ionization potential:
i.e. the binding energy of the $2p$ electron: 21.6~eV.
If instead we use the binding energy
of the $2s$ electron then we predict a cutoff of 
\begin{equation*}
  E_{\mathrm{cutoff,2s}} \approx 3.2U_p+E_{2s} =  160~\mathrm{eV}, 
\end{equation*}
in line with the 45~eV XUV spectrum in Fig.
\ref{fig:short_long}. 

In order to isolate the contribution of the $2s$ electron two
calculations are performed with an XUV photon energy of 45~eV: one comprising both the $2s^22p^5$ and $2s2p^6$
Ne$^+$ thresholds, and one with only the $2s^22p^5$. Removing the $2s2p^6$
threshold effectively neglects ionization of the $2s$ electron in the
calculation. A similar approach was used to elucidate the contribution of the
$3s$ and $3p$ electrons to low energy harmonic generation in Ar \cite{brown_prl}.

\begin{figure}[t]
\begin{centering}
\includegraphics[width=7.5cm]{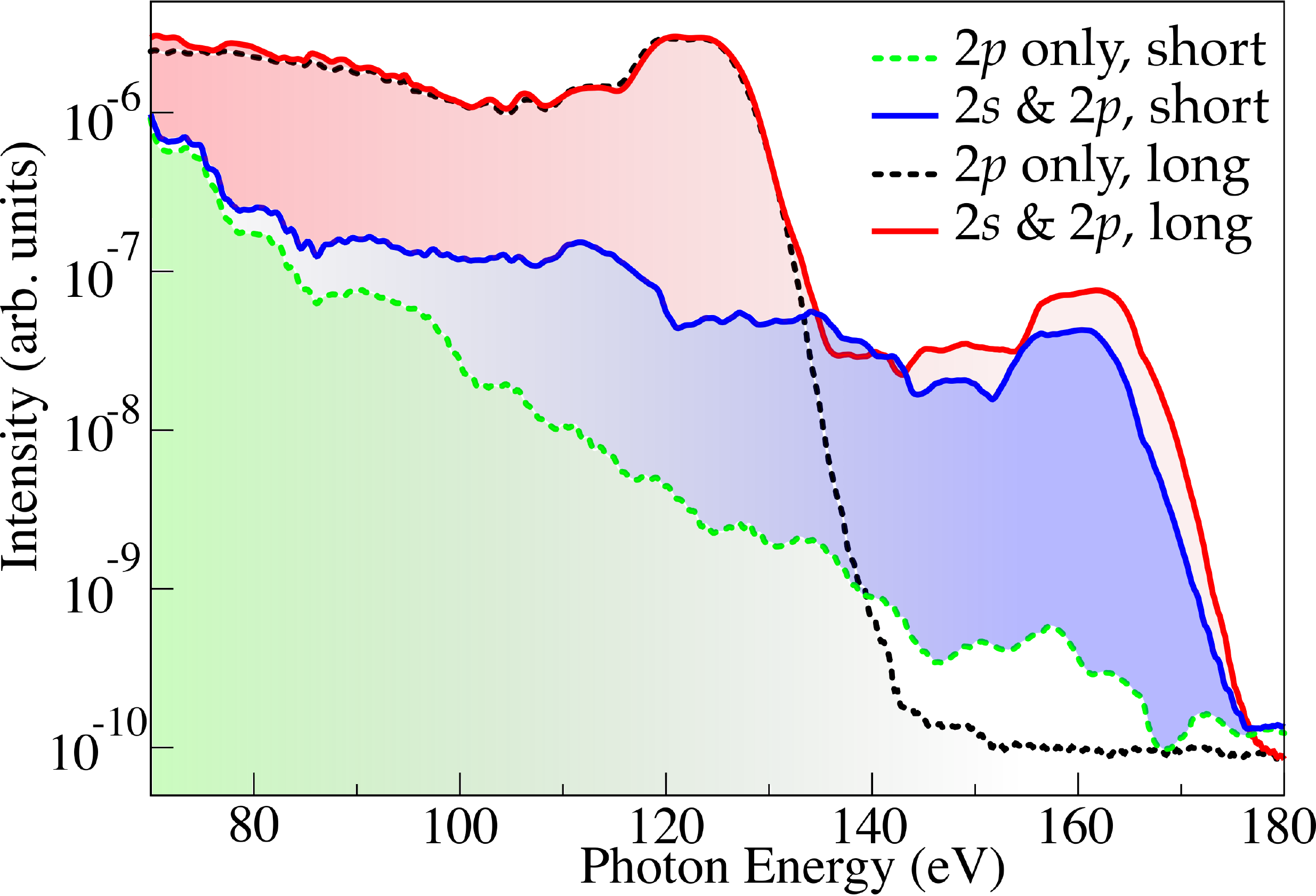}
\caption{(Color online) 
The smoothed cutoff harmonics from Ne driven by an \fs{18},
\um{1.8}, \intensity{1.2}{14} pulse, in
combination with a \as{250} (lower two lines), and \fs{1} (upper two lines),
45~eV XUV pulse timed to coincide with the penultimate maximum of the IR field. Calculations include the action of both the $2s$ and $2p$
electrons (solid lines) or neglecting the action of the $2s$ (dotted lines).
\label{fig:cutoff}}
\end{centering}
\end{figure}

Figure \ref{fig:cutoff} shows the high energy harmonics from Ne as generated with 
these two atomic structure configurations, for both the long and short XUV pulses.
There is a clear difference between the spectra from the $2p-$only and
$2s-2p$ calculations: the $2s-2p$ yield shows a recognisable
plateau extending to a cutoff at 160~eV which is not present in the $2p-$only
calculations for either long or short pulses. This confirms that the
highest energy harmonics are generated by the action of the $2s$ electron. 
%Again we note
%that a photon energy of 45~eV is not sufficient to photoionize the $2s$ electron
%directly, but a two-stage ionization process occurs involving photoexcitation to
%the $2s2p^6{}np$ Rydberg states and subsequent field ionization. This is again
%supported by calculations which show the same extension in cutoff for
%time-delays between 2.5 and \fs{3.5} (Fig. \ref{fig:td_scan}).

In the long pulse, there is no difference between the $2p-$only and $2s-2p$
spectra up to an
initial cutoff at 125~eV.
For the short pulse, the overall harmonic yield is largely suppressed, but
the integrated yield above 100~eV is increased by an order
of magnitude by the inclusion of the $2s$ electron. This is surprising as the
ionization probability for a $2s$ electron is around 1\% of that of a
$2p$ electron. Furthermore, the cutoff ascribed to the $2p$ electron-- 125eV--
does not agree with the classically predicted cutoff energy of 137~eV. 

To
elucidate these differences we perform classical trajectory simulations: for each
`birth time', we can calculate the electron's velocity and position by
integrating over the acceleration in the IR field. Those 
trajectories which pass again through the origin describe recolliding electrons
from which the recollision energy-- and hence the energy of the HHG radiation-- can be
determined. Importantly, in general, a photoionized electron will be born
into the field with non-zero velocity, while in the traditional three-step model
the initial velocity of the tunnel-ionized electron is assumed to be zero
\cite{path_control_efficiency}. This
initial velocity is calculated directly from the excess energy absorbed by the
electron: i.e. the difference between the XUV-photon energy and the ionization
potential. We note that this velocity may be in any direction relative to
the linearly polarized IR field. Thus a $2p$ electron will gain an excess energy of 23.4~eV
($45-21.6$~eV), while a $2s$ electron is born with zero initial velocity (in
practice there will be a spread of initial energies due to the bandwidth of the
XUV pulse, but we assume just one initial velocity for the purposes of the
model). 

\begin{figure}[t]
\begin{centering}
\includegraphics[width=7.5cm]{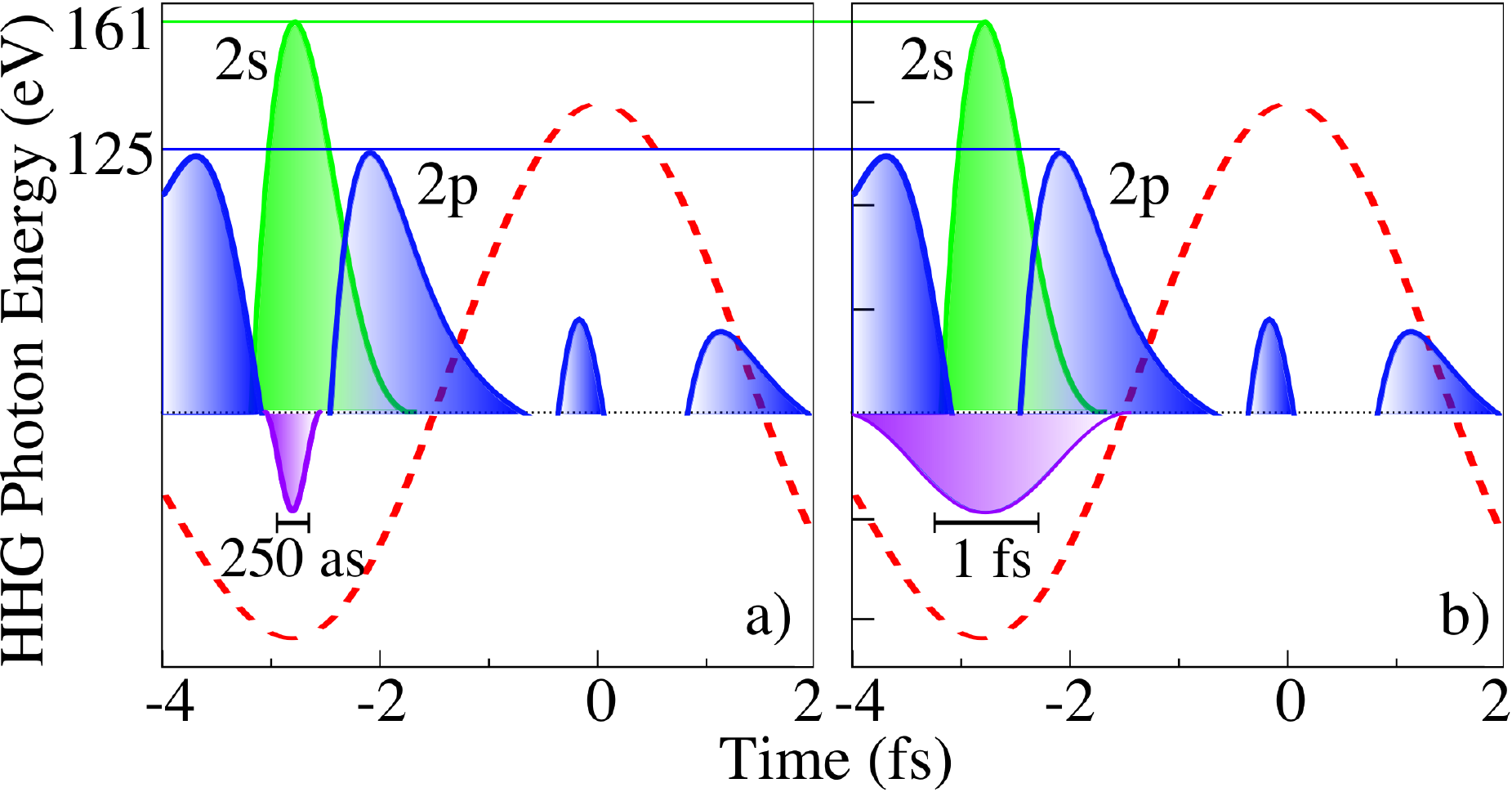}
\caption{(Color online) The maximum emitted harmonic photon energies that would result from electrons being
  released at the times shown along the horizontal axis 
  for both $2p$ (blue) and $2s$ (green)
  electrons from Ne in a \fs{18},
\um{1.8}, \intensity{1.2}{14} pulse (red dashed line). 
Trajectories are initiated by a a) \as{250} and b) \fs{1} 45~eV XUV pulse. 
Trajectories of $2s$ and $2p$ which lead to recollision  are localized
in time, and therefore can be resolved in the short-pulse case (a).
For a long XUV pulse (b), both electrons contribute.
  \label{fig:windows}}
\end{centering}
\end{figure}

Figure \ref{fig:windows} shows the HHG photon energy for $2s$ and $2p$ electron
recollision trajectories originating at different times in the IR pulse. The excess energy gained 
negates the recollision of $2p$ electrons born at the penultimate peak of the IR, but allows
those born earlier/later in the IR cycle to recollide. By contrast, the choice
of an XUV-photon energy below the $2s$ ionization threshold ensures that the 
$2s$ electron is born with zero velocity. Thus its recollision trajectories
correspond to birthtimes around the penultimate peak of the IR field. 
As tunnel ionization due only to the IR field is negligible, we assume
that only trajectories born during the XUV pulse contribute substantially to the HHG
spectrum and-- in the short-pulse case-- this means only $2s$ electrons will
contribute. This is despite the
relatively low probability of ionizing a $2s$ electron versus a $2p$ electron.

By contrast, the long pulse does not resolve the contribution of the $2s$ and
$2p$ electrons. Thus both contributions are present. 
However, because the ionization probability for a $2p$ electron is two orders of
magnitude larger than for a $2s$, the HHG yield up to the first cutoff is
dominated by $2p$ electrons, and we only elucidate the $2s$ contribution above
this energy.  The initial velocity of the $2p$ electrons in the field modifies
their trajectory from the tunnel-ionization case, reducing the maximum energy of
the recolliding electron, and hence the initial $2p$ cutoff of the HHG spectrum.
The maximum energy calculated for the $2p$ electron is 125~eV, in
line with the $2p$ cutoff observed in the spectra (Fig.
\ref{fig:cutoff}). 

\begin{figure}[t]
\begin{centering}
\includegraphics[width=7.5cm]{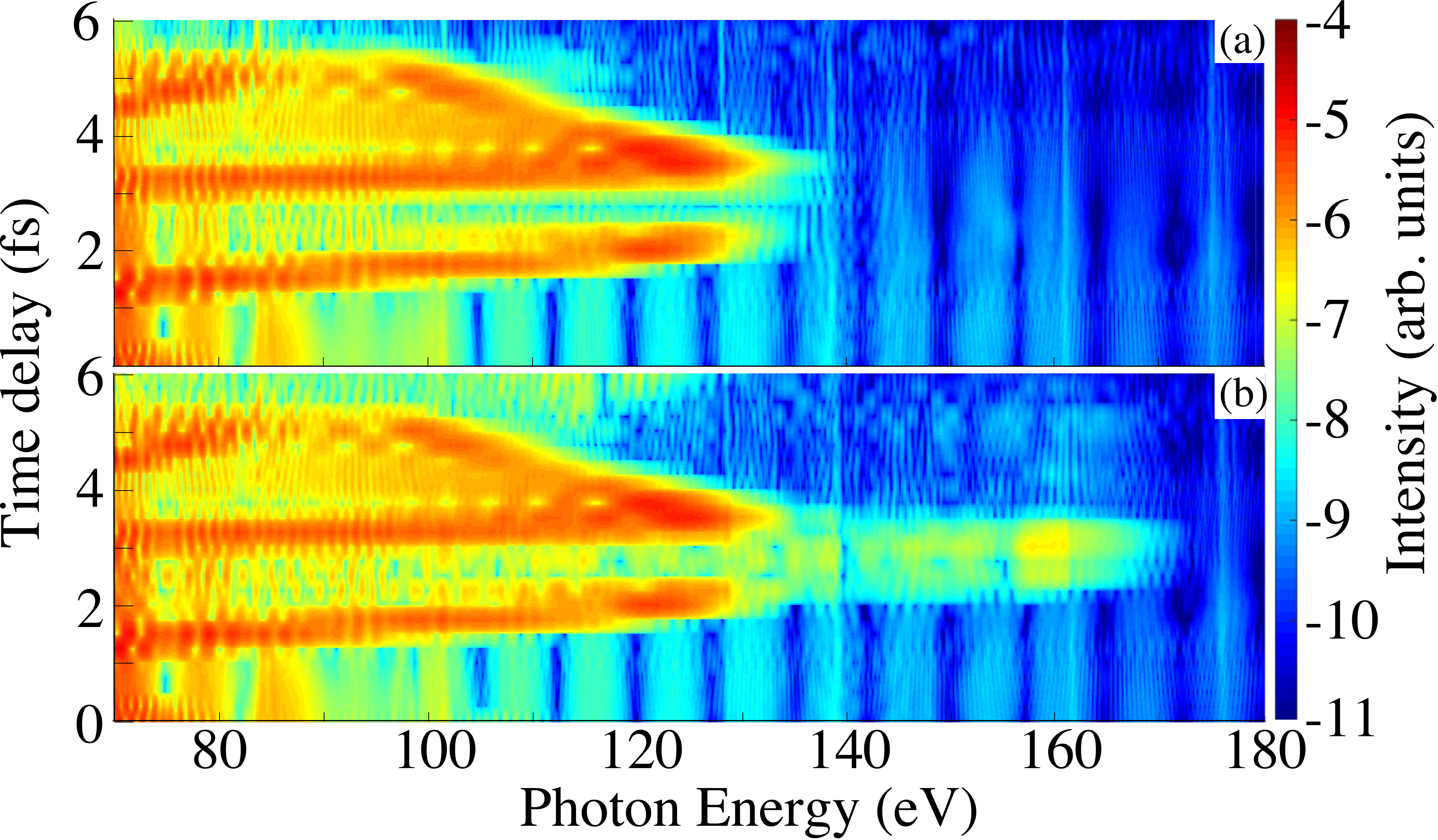}
\caption{(Color online) The XIHHG spectrum for Ne as a function of time delay
  between the XUV and IR pulse peaks for RMT calculations both
  (a) neglecting and (b) including the action of a $2s$ electron.
  \label{fig:td_scan}}
\end{centering}
\end{figure}

In order to confirm the intuition of this simple model we perform calculations
for the XUV-photon energies shown in Fig. \ref{fig:short_long}. The yellow
squares shown are the predicted cutoff energies for the $2p$ trajectories as
calculated by the model, while orange circle and cross are those calculated
for the $2s$ electron. They are seen to be in good agreement with the
RMT results. 

% probably not necessary
%We note that for higher XUV energies, these trajectories 
%can be entirely suppressed, because the IR field is insufficiently
%strong to reverse the outgoing velocity of the electron. In particular, if the
%initial pulse is tuned to ionize the core, $1s$ electron (870~eV), then no
%trajectories persist for either the $2s$ or $2p$ electrons. Thus, XIHHG can
%be selectively sensitive only to the contribution of core electrons if the
%ionizing photon is sufficiently energetic.
%%%%%%%%%%%%%

To confirm the two-stage ionization mechanism, we refer to
Fig. \ref{fig:td_scan}, which shows the time-delay scans for
two simulations for the \as{250}, 45~eV XUV pulse for calculations
including or neglecting the action of the $2s$ electron. The time window in which
the $2p$ recollision is suppressed is clearly seen between 2.5 and \fs{3.5}.
This corresponds with a drop in overall yield, but an extension of the cutoff
energy with the contribution of the $2s$ electron. The cutoff energy is in line
with the prediction of the model calculations shown in Fig.
\ref{fig:windows}. In particular we observe enhancement at the $2p$ cutoff for
time delays around \fs{2} and \fs{4} which match the predictions of the model
(The cutoff energy observed is higher than in the model calculations but within
the bandwidth of the XUV pulse, which is not accounted for in the model).

Importantly, the extension of the cutoff due to the $2s$ electron is observed
for time delays between 2.25 and \fs{3.5}, corresponding to the XUV pulse
arriving at, or just before, the IR peak. This confirms the proposed two-stage ionization
mechanism, wherein the $2s$ electron is promoted into a $2s2p^5$$n\ell$ Rydberg
state before being field ionized at the IR peak. This mechanism warrants further
investigation as a scheme may be devised to probe sensitively the lifetimes of excited
Rydberg states.

Because the binding energy of inner-valence electrons is not well separated from
the ionization potential of the valence shell, XIHHG of inner-valence electrons
will not provide substantial gains in cutoff energy. This could be accomplished
more easily with the use of higher intensity IR pulses, or indeed with XIHHG of
more deeply bound electrons. However, the scheme still represents more promise
than may previously have been thought. Due to the relatively small probability of
ionizing an inner-valence electron-- even with an XUV pulse tuned to the correct
binding energy-- the outer-valence electron should dominate the HHG
process, and indeed, for long XUV pulses we observe that the yield from
neon is dominated by the $2p$ electron contribution up to the $2p$ cutoff. However,
we have shown that with short pulses, whose energy is below the $2s$ ionization
energy, it is possible to selectively
suppress the contribution of $2p$ electrons, and elucidate the
contribution of the $2s$ electron, even if this results in a much lower overall
yield. 

Thus for attosecond duration XUV pulses delayed appropriately with respect to
the IR field, it should be possible to perform HHG spectroscopy of inner-valence
electrons.  If appropriate pulse profiles, or as-pulse trains 
can be devised both inner- and outer-valence electrons may
contribute on a similar magnitude.
This would then facilitate HHG spectroscopy of interference dynamics as
previously explored for low energy harmonics \cite{brown_prl}.

%Thus, we have shown that by driving HHG with an attosecond XUV pulse it is
%possible to selectively enhance and suppress the yield due to particular
%electrons. This is important for inner valence electrons in
%particular, where the relative ionization probabilities would otherwise lead to
%the signal from the inner electron being masked by that from the outer. It is
%expected that the ability to selectively drive HHG from particular electrons
%should enable sensitive spectroscopic measurements of deeper lying states in
%atoms and molecules. This has already been shown to be important for
%understanding the dynamics of core-hole states
%\cite{correlation_core_hole_dynamics}. 

%\begin{acknowledgments}
HWH acknowledges
financial support from the UK EPSRC under grant no. EP/G055416/1 and the EU Initial
Training Network CORINF. 
This work used the
ARCHER UK National Supercomputing Service (\url{http://www.archer.ac.uk}).
The data used in this paper may be accessed at
Ref. \cite{xuv_hhg_data}.
%\end{acknowledgments}

\bibliography{/users/abrown41/Documents/pub/bib_utils/mybib}

\end{document}